\documentclass[
 preprint,amsmath,amssymb,
 aps
]{revtex4-2}

\usepackage{graphicx}
\usepackage{mhchem}
\usepackage{amsmath,amssymb}
\usepackage{xcolor}
\usepackage{siunitx}

\begin{document}

\title{Path-integral treatment of charged Bose polarons}

\author{Laurent H. A. Simons}
\affiliation{
 Theory of Quantum and Complex Systems, Physics Department, Universiteit Antwerpen, B-2000 Antwerpen, Belgium
}

\author{Michiel Wouters}
\affiliation{
 Theory of Quantum and Complex Systems, Physics Department, Universiteit Antwerpen, B-2000 Antwerpen, Belgium
}

\author{Jacques Tempere}
\affiliation{
 Theory of Quantum and Complex Systems, Physics Department, Universiteit Antwerpen, B-2000 Antwerpen, Belgium
}

\date{\today}

\begin{abstract}
The system of a charged impurity in an interacting Bose gas has gained significant attention due to the long-range ion-atom interactions and the study of transport properties. Here, the ground state energy of a charged Bose polaron is calculated within the Bogoliubov approximation for both the Fröhlich and beyond-Fröhlich Hamiltonians using a generalized Feynman variational path-integral approach, which obtained accurate results for other polaron problems. The generalized approach, which was used to improve the energy result for the neutral polaron, has resulted in a minor improvement, indicating that Feynman's approach is sufficient when the impurity-boson interaction is long-range. Beyond-Fröhlich corrections results in the emergence of a divergence in the polaronic energy indicating a transition between the repulsive and attractive polaron regime. The path-integral approach with the beyond-Fröhlich Hamiltonian is also compared to a field-theory calculation from Christensen et al, 2021. The validity of the Bogoliubov approximation is investigated. The optical absorption has also been calculated within the Bogoliubov approximation for weak ion-atom interactions, and the effect of finite temperature has been studied. We show that the coupling of the ion to an oscillating external electric field offers a straigtforward experimental probe for the charged polaron in a Bose gas, different from but complementary to existing spectroscopic techniques.
\end{abstract}

\maketitle
\newpage

\section{\label{sec:introduction}Introduction}
A charge inserted into an ionic crystal displaces the ions inside the medium resulting in an induced polarization cloud that will follow the charge as it moves throughout the medium \cite{devreese2016fr,franchini2021polarons}. The charge, in conjunction with the induced polarization cloud, can be described as a quasiparticle, which was called a \textit{polaron} by Pekar \cite{pekar1946local}. The polaron concept has since then been generalized to any system consisting of a particle interacting with a generic bath of bosons via emission or absorption of a boson from the bath \cite{devreese2016fr,franchini2021polarons}. 
A recent example is the Bose polaron, formed when an impurity is introduced in a Bose-Einstein condensate (BEC) and the impurity interacts with the bosonic bath of excitations of the condensate, the Bogoliubov modes.  
The Bose polaron has been realized experimentally \cite{arlt2016expt,jin2016expt,yan2020bose,arlt2022lifetime} and has been studied extensively in the literature using an array of different analytical and numerical methods \cite{tempere2009feynman,grusdt2015renormalization,ardila2015qmc,vlietinck2015diagrammatic,shchadilova2016quantum,shchadilova2016corgauss,levinsen2017temp,ichmoukhamedov2019feynman}. 
However, most of the studies limit themselves to neutral impurities where the impurity-boson interaction is taken to be a contact interaction. Other impurity-boson interactions can be studied, such as a dipolar interaction \cite{ardila2018ground,kain2014polarons,sanchez2023universal,volosniev2023non} or a long-range atom-ion interaction \cite{casteels2011polaronic,christensen2021charged,astrakharchik2022charged, astrakharchik2021ionic}. 
The system of an ion in a BEC is of interest due to the possibility of probing the ion using an electric field and the long-range nature of the atom-impurity interaction. The absorption spectrum has already been studied and calculated for the solid-state polaron \cite{feynman1962mobility,devreese1972optical}, however it is not possible to study it for the neutral Bose polaron as the impurity is neutral.
The problem of a charged impurity in a condensate was studied in Refs.~\cite{astrakharchik2021ionic, astrakharchik2022charged} using Monte Carlo techniques, while in Ref.~\cite{christensen2021charged} the polaronic energy was calculated using the Bethe-Salpeter equation with the ladder approximation. 

In this paper, the polaronic energy and optical absorption of charged Bose polarons are calculated. The Feynman variational path-integral approach \cite{feynman1955slow} is used to determine the energy of the charged Bose polaron by using a variational trial action yielding an upper bound on the free energy. 
It has been applied to various problems in the past and has been found to produce accurate results comparable to those obtained from exact Monte-Carlo methods while being numerically less demanding \cite{vlietinck2015diagrammatic,ichmoukhamedov2019feynman,ichmoukhamedov2022general}. 
It has been used before for the charged Bose polaron in Ref. \cite{casteels2011polaronic}, however, here, a more general quadratic trial action is used \cite{rosenfelder2001best,ichmoukhamedov2022general} and so-called beyond-Fröhlich terms \cite{shchadilova2016quantum,ichmoukhamedov2019feynman} are included, which are particularly important for strong ion-atom interactions. 
The result for the polaronic energy is also compared to the mean-field result and the result from Ref. \cite{christensen2021charged} using the ladder approximation. A small improvement is observed compared to the result from Ref. \cite{christensen2021charged}. In addition, we calculate the number of phonons and the polaron radius. This allows to determine the local depletion of the condensate, and delineate the regime where the Bogoliubov approximation remains valid.

The second part of this paper focuses on absorption spectroscopy for charged Bose polarons. Neutral Bose polarons cannot be probed by direct light absorption, and one has to rely on Bragg spectroscopy \cite{casteels2011bragg} or on spectroscopically measuring the energy required to transfer a few atoms into the impurity state \cite{arlt2016expt,jin2016expt,yan2020bose,arlt2022lifetime}. For charged Bose polarons, we propose absorption spectroscopy as a new way to probe the polaronic state. For solid-state polarons, the optical absorption has proven to be a powerful technique to assert the presence of polarons and extract their properties. Here, we compute the  absorption spectrum in the weak interaction regime, where it is assumed that there is no phonon entanglement, using a method based on Ref.~\cite{houtput2022optical}. This spectrum can be revealed experimentally by exposing the condensate with its immersed ion to an AC electric field (in a frequency range around \SI{3}{\kilo\hertz} for the system of a \ce{^{87}_{} Rb^+} ion in a \ce{^{87}_{} Rb} condensate). The absorbed energy as a function of frequency can be detected as attenuation (via the complex impedance) of the applied AC electric field, or alternatively as heating of the condensate.

This paper is organised as follows. In Section II, the Hamiltonian is derived for the charged Bose polaron within the Bogoliubov approximation. In Section III, the polaronic energy is calculated for (A) the Fr\"ohlich Hamiltonian and (B) the full Bogoliubov Hamiltonian including beyond-Fr\"ohlich corrections using the general memory kernel approach. The latter result is also compared to a result from Ref. \cite{christensen2021charged}. The validity of the Bogoliubov approximation is checked in Section V by calculating the phonon number and the polaron radius. In Section VI, the optical absorption is calculated within the Fr\"ohlich model for weak interaction where there is no phonon entanglement. In the last section, Section VII, the conclusion is given, and possible next steps are discussed.

\section{\label{sec:hamiltonian}Hamiltonian}
The general Hamiltonian describing an impurity with mass $m_I$ in a gas of interacting bosons with mass $m_B$ is given by \cite{tempere2009feynman}
\begin{align}
    \hat{H} &= \frac{\hat{\mathbf{p}}^2}{2m_I}+\sum_{\mathbf{q}}\epsilon_{\mathbf{q}} \hat{a}^\dagger_{\mathbf{q}}\hat{a}_{\mathbf{q}} \nonumber\\
    &+ \frac{1}{2}\sum_{\mathbf{k},\mathbf{k}',\mathbf{q}}V_{BB}(\mathbf{q})\hat{a}^\dagger_{\mathbf{k}'-\mathbf{q}}\hat{a}^\dagger_{\mathbf{k}+\mathbf{q}}\hat{a}_{\mathbf{k}}\hat{a}_{\mathbf{k}'} \nonumber\\
    &+\sum_{\mathbf{k},\mathbf{q}}V_{IB}(\mathbf{q})\hat{\rho}_I(\mathbf{q})\hat{a}^\dagger_{\mathbf{k}-\mathbf{q}}\hat{a}_{\mathbf{k}},
\end{align}
with $\epsilon_{\mathbf{q}} = (\hbar q)^2/(2m_B)$.
The first two terms represent the free impurity and a free gas of bosons, respectively. $\hat{a}^\dagger_{\mathbf{q}}, \hat{a}_{\mathbf{q}}$ represent the creation and annihilation operators of a boson with momentum $\mathbf{q}$. The third term represents the interaction between the bosons 
characterized by an interaction amplitude related to the Fourier transform of the boson-boson interaction potential which is assumed to be a contact potential $V_{BB}(\mathbf{q})=g_{BB}/V=4\pi\hbar^2a_s/(m_BV)$ with $a_s$ the scattering length and $V$ the condensate volume. The interaction between the impurity and the bosons is described by the last term, where 
$\hat{\rho}_I(\mathbf{q})=\exp(i\mathbf{q}\cdot\hat{\mathbf{r}})$ is the impurity density and $V_{IB}(\mathbf{q})$ is the impurity-boson interaction potential in momentum space.
As has been done in the case of a neutral impurity \cite{tempere2009feynman}, the Bogoliubov prescription \cite{bogoliubov1947theory} can be applied. The Bogoliubov approximation states that for weak boson-boson interactions $a_sn^{1/3}\ll1$, with $n$ the density, the creation and annihilation operators for momentum zero can be replaced by the c-number $\sqrt{N_0}$. Then, the Bogoliubov transformation is used to diagonalize the quadratic part of the interacting Bose gas term of the Hamiltonian by introducing new Bogoliubov operators $\hat{\alpha}^\dagger_{\mathbf{q}},\hat{\alpha}_{\mathbf{q}}$. 
The resulting Hamiltonian is similar to the neutral impurity case; however, there is a general impurity-boson interaction instead of a contact interaction \cite{shchadilova2016quantum,ichmoukhamedov2019feynman}
\begin{align}
    \hat{H} = &  \frac{\hat{\mathbf{p}}^2}{2m_I}+\sum_{\mathbf{k}}\hbar\omega_{\mathbf{k}}\hat{\alpha}^\dagger_{\mathbf{k}}\hat{\alpha}_{\mathbf{k}} +\sqrt{N_0} \sum_{\mathbf{k}} V_{IB}(\mathbf{k})\hat{\rho}_I(\mathbf{k})V_{\mathbf{k}}(\hat{\alpha}_{\mathbf{k}}+\hat{\alpha}^\dagger_{-\mathbf{k}}) 
    \nonumber \\ & 
    + \sum_{\mathbf{s}\neq0}\sum_{\mathbf{k}\neq0}V_{IB}(\mathbf{k}-\mathbf{s})\hat{\rho}_I(\mathbf{k}-\mathbf{s})W^{(1)}_{\mathbf{k},\mathbf{s}}\hat{\alpha}^\dagger_{\mathbf{s}}\hat{\alpha}_{\mathbf{k}} \nonumber \\ 
    & +  \frac{1}{2}\sum_{\mathbf{s}\neq0}\sum_{\mathbf{k}\neq0}V_{IB}(\mathbf{k}-\mathbf{s})\hat{\rho}_I(\mathbf{k}-\mathbf{s})W^{(2)}_{\mathbf{k},\mathbf{s}}(\hat{\alpha}^\dagger_{\mathbf{s}}\hat{\alpha}^\dagger_{-\mathbf{k}}+\hat{\alpha}_{-\mathbf{s}}\hat{\alpha}_{\mathbf{k}}),
\end{align}
with $\hbar\omega_{\mathbf{k}}=\hbar^2k/(2m_B\xi)\sqrt{2+(\xi k)^2}$ the Bogoliubov dispersion, $V_{\mathbf{k}}=((\xi k)^2/(2+(\xi k)^2))^{1/4}$, $W^{(1,2)}_{\mathbf{k},\mathbf{s}}=(V_{\mathbf{k}}V_{\mathbf{s}}\pm V^{-1}_{\mathbf{k}}V^{-1}_{\mathbf{s}})/2$, and the healing length $\xi = 1/\sqrt{8\pi a_s n}$.
The third term describes a Fröhlich-like interaction between the impurity and the Bogoliubov phonons, where the impurity absorbs or emits a Bogoliubov phonon. However, compared to the simple Fröhlich model \cite{frohlich1954electrons} of a solid-state polaron, the Hamiltonian also includes additional terms with two creation and annihilation operators, which will be referred to as beyond-Fröhlich corrections \cite{shchadilova2016quantum,ichmoukhamedov2019feynman}. They are only important for strong ion-atom interactions.
The charged impurity has a long-range interaction potential compared to the neutral impurity system. The following pseudopotential is used \cite{astrakharchik2021ionic,christensen2021charged}:
\begin{equation}
    V_{IB}(r)=-\frac{a}{(r^2+b^2)^2}\frac{r^2-c^2}{r^2+c^2},
\end{equation}
where $a$ characterizes the strength of the ion-atom interaction potential, while $b$ is related to the depth of the potential. The last term, depending on $c$, creates a repulsive barrier in the potential for $r<c$. The Fourier transform of the above potential is given by
\begin{equation}
    V_{IB}(q) = \frac{a\pi^2}{bV}\frac{4bc^2(e^{-cq}-e^{-bq})-q(b^4-c^4)e^{-bq}}{(b^2-c^2)^2q}.
\end{equation}
The parameters $a$, $b$, and $c$ can be linked to the atom-ion scattering length \cite{casteels2011polaronic, christensen2021charged}. 
Compared to the neutral impurity problem, the potential converges to zero for large momentum. This means that in the following calculations, no cutoff is needed for the momentum integrals. 
A dimensionless coupling constant can be introduced similarly to the solid-state polaron and neutral Bose polaron. It is defined as \cite{casteels2011polaronic}
\begin{equation}
    \alpha = \frac{R^4}{a_sb^2\xi}, \quad R=\sqrt{2m_r a/\hbar^2},
\end{equation}
where $m_r=m_Bm_I/(m_B+m_I)$ is the reduced mass.
In the next section, the ground state energy of the charged Bose polaron is calculated using the variational path integral method (a) within the Fröhlich approximation and (b) including the beyond-Fröhlich terms.

\section{\label{sec:energy}Polaronic energy}
The method used in this section is based on the Feynman-Jensen inequality, which states that an upper bound for the free energy of a system with action $S$ can be found by introducing an exactly solvable variational trial action $S_0$ and minimizing for the variational parameters
\begin{equation}
    F \leq F_0 + \frac{1}{\hbar\beta}\langle S-S_0\rangle_0,
\end{equation}
where $F_0$ is the free energy of the trial action and $\beta=1/(k_BT)$ the inverse temperature. In this paper, the following general exactly solvable quadratic trial action is used
\begin{equation}
    S_0=\int \limits_0^{\hbar\beta} \frac{m_I}{2}\dot{\mathbf{r}}^2(\tau)d\tau +\frac{m_I}{2\hbar\beta}\int\limits_0^{\hbar\beta}\int\limits_0^{\hbar\beta}x(\tau-\sigma)\mathbf{r}(\tau)\cdot\mathbf{r}(\sigma)d\tau d\sigma,
\end{equation}
where $x(\tau-\sigma)$ is a variational memory kernel \cite{rosenfelder2001best,ichmoukhamedov2022general}.
There are two restrictions on the kernel: it is periodic $x(\hbar\beta-\tau)=x(\tau)$ and $\int_0^{\hbar\beta} x(\tau)d\tau=0$ \cite{ichmoukhamedov2022general}. In the free energy, the kernel in frequency space is used defined by $x(\nu)=1/(\hbar\beta)\int_0^{\hbar\beta} x(\tau)e^{-i\nu\tau} d\tau$. The free energy and expectation value of the trial action can be found in the literature \cite{ichmoukhamedov2022general}.

\subsection{Fröhlich approximation}
In this section, the polaronic energy of the charged Bose polaron is calculated using the derived Hamiltonian without the beyond-Fröhlich terms.

A Lagrangian (and corresponding action functional) can be derived corresponding to the Bogoliubov-Fr\"ohlich Hamiltonian \cite{houtput2022voorbij}. However, the corresponding action depends on the electron \textit{and} phonon degrees of freedom. The phonon degrees of freedom can be removed by integrating them out, as the path integral of a forced harmonic oscillator can be solved exactly. The effective action of the Fr\"ohlich Hamiltonian is given by \cite{tempere2009feynman}
\begin{align}
    S &=\int \limits_0^{\hbar\beta} \frac{m_I}{2}\dot{\mathbf{r}}^2(\tau)d\tau -\frac{N_0}{2\hbar}\sum_{\mathbf{k}}V^2_{IB}(\mathbf{k})V^2_{\mathbf{k}} \nonumber \\
    &\times \int \limits_0^{\hbar\beta}\int \limits_0^{\hbar\beta}\frac{\cosh[\omega_{\mathbf{k}}(|\tau-\sigma|-\hbar\beta/2)]}{\sinh(\omega_{\mathbf{k}}\hbar\beta/2)}e^{i\mathbf{k}\cdot(\mathbf{r}(\tau)-\mathbf{r}(\sigma))}d\tau d\sigma.
\end{align}
The free energy for a neutral Bose polaron has been calculated using the aforementioned trial action \cite{ichmoukhamedov2022general} and can be adjusted to include a long-range interaction potential giving
\begin{align}
    F\leq & \frac{3\hbar}{2\pi}\int_0^\infty d\nu \left(\log\left(1+\frac{x(\nu)}{\nu^2}\right)-\frac{x(\nu)}{x(\nu)+\nu^2}\right) \nonumber \\
    & -\frac{N_0V}{2\pi^2\hbar}\int_0^\infty dk k^2V^2_{IB}(k)V^2_k\int_0^{\infty}due^{-\omega_{\mathbf{k}}u}\mathcal{F}_k(u),
    \label{energy}
\end{align}
where
\begin{align}
    x(\nu)= & \frac{2N_0V}{3m_I\pi^2\hbar}\int_0^\infty dk k^4V^2_{IB}(k)V^2_k \nonumber\\
    & \times \int_0^{\infty}du \sin^2(\nu u/2) e^{-\omega_{\mathbf{k}}u}\mathcal{F}_k(u),
\end{align}
with
\begin{equation}
    \mathcal{F}_k(u)=\exp\left(-\frac{\hbar k^2}{m_I\pi}\int_0^\infty d\nu \frac{1-\cos(\nu u)}{x(\nu)+\nu^2}\right).
\end{equation}
The equation for the kernel can be found by taking the functional derivative of the free energy with respect to the kernel and setting it equal to zero \cite{ichmoukhamedov2022general}.
In the above equations, the temperature is set to zero to simplify calculations.
The kernel needs to be calculated iteratively. The initial ansatz for the kernel is the Lee-Low-Pines kernel given by $x(\nu)=0$. The relative tolerance chosen for the energy convergence is $10^{-3}$. 
Gauss-Legendre quadrature with $20$ points with $200$ subintervals and a substitution resulting in a finite integration range was used based on Ref. \cite{rosenfelder2001best} to calculate the integrals and resulted in convergence for the polaronic energy.

\begin{figure}
    \centering
    \includegraphics[scale=1]{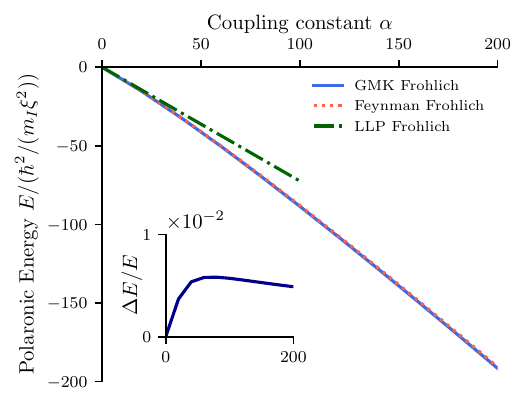}
    \caption{The free energy of the charged Bose polaron using the Fr\"ohlich model for $b=0.1\xi,c=0.0023\xi,m_B=m_I$ as a function of the coupling strength $\alpha$. The blue line shows the result using the general memory kernel (GMK) method, while the red dotted line shows the result calculated using the original Feynman kernel as \cite{casteels2011polaronic}. The green dash-dotted line shows the weak-coupling limit Lee-Low-Pines energy. The inset shows the relative difference $\Delta E/E$ between the GMK and Feynman energy; the difference is below 1 percent.}
    \label{fig1}
\end{figure}

The second term of the free energy (\ref{energy}) is proportional to the coupling constant $\alpha$ defined before, similar to the neutral Bose polaron. The free energy was calculated for values of the coupling constant ranging from $0$ to $200$, which are realistic as the scattering length is two orders of magnitude smaller than the healing length and $R$. $c=0.0023\xi$ is chosen using Ref. \cite{christensen2021charged}, $b$ is fixed to $0.1\xi$, and $R=\xi$. The mass of the impurity is also set equal to the mass of the bosons like in the \ce{^{87}_{} Rb^+}-\ce{^{87}_{} Rb} system used in Ref. \cite{christensen2021charged}. The result is shown in Figure \ref{fig1}. Compared to the neutral Bose polaron problem \cite{tempere2009feynman,ichmoukhamedov2022general}, the free energy is always negative within the Fröhlich model.
The result of the free energy is also compared to the results of two other methods: the Lee-Low-Pines approach, which is only valid for weak coupling strengths, and the Feynman variational method with the original Feynman kernel, which was applied to the charged Bose polaron problem in Ref. \cite{casteels2011polaronic}.

The Lee-Low-Pines result \cite{lee1953motion, grusdt2015new} can be derived by applying a unitary transformation removing the electron degrees of freedom and by using a coherent state variational ansatz. The resulting equation for the Lee-Low-Pines energy of the charged Bose polaron is
\begin{equation}
    E_{LLP}=-\frac{N_0V}{2\pi^2\hbar}\int_0^\infty dk \frac{k^2V^2_{IB}(k)V^2_k}{\omega_k+\hbar k^2/(2m_I)}.
\end{equation}
The above result can also be found by placing the kernel equal to zero in the general memory kernel energy equation (\ref{energy}). The Lee-Low-Pines energy is linearly proportional to the coupling strength and is only correct for small $\alpha$. The Lee-Low-Pines energy is plotted in Figure \ref{fig1}.
The original Feynman trial action used by \cite{casteels2011polaronic} for the charged Bose polaron is given by \cite{tempere2009feynman}
\begin{align}
    S_0= & \int_0^{\hbar\beta} \frac{m_I}{2}\dot{\mathbf{r}}^2(\tau)d\tau +\frac{MW^3}{8} \nonumber \\
    & \times \int_0^{\hbar\beta} d\tau \int_0^{\hbar\beta} d\sigma 
    \frac{\cosh(W(|\tau-\sigma|-\hbar\beta/2))}{\sinh(W\hbar\beta/2)}(\mathbf{r}(\tau)-\mathbf{r}(\sigma))^2.
\end{align}
It depends on two variational parameters $M$ and $W$. An equation can be derived for the free energy as a function of the variational parameters \cite{casteels2011polaronic}. The variational parameters are found by minimizing the energy. The energy obtained using the original Feynman kernel is plotted in Figure \ref{fig1}.
Compared to the neutral Bose polaron problem, the improvement of the energy result by using the general memory kernel approach instead of the original Feynman approach is smaller. This was also found to be the case for the solid-state polaron \cite{rosenfelder2001best}. This shows that the original Feynman kernel works well for finite-range potentials for which there is no UV divergence in the momentum integral in (\ref{energy}). If a contact interaction is used, the general memory kernel approach must be used to obtain correct results for the polaronic energy. For the values chosen for Figure \ref{fig1}, the improvement is always less than 1 percent, as seen in the inset. While they give similar energies, the kernels are still different, so the above conclusions can not be extended to other polaron properties, such as the effective mass.
However, for large coupling strengths, the above results do not hold, as the beyond-Fröhlich terms are no longer negligible. In the next subsection, the path integral method is adjusted to accommodate the beyond-Fröhlich terms following \cite{ichmoukhamedov2019feynman}.

\subsection{\label{sec:beyondfrohlich}Including beyond-Fr\"ohlich corrections}
For the neutral Bose polarons, beyond-Fröhlich corrections were studied within the path-integral formalism in Ref.~\cite{ichmoukhamedov2019feynman}. 
A Lagrangian was set up for the beyond-Fröhlich Hamiltonian, and the effective action was derived by introducing source terms and a perturbative expansion. 
However, during the derivation of the Lagrangian, there is a constraint: the potential needs to be separable. 
This is not a problem in the case of a contact potential, as for the neutral impurity, however here the condition $V_{IB}(\mathbf{k}-\mathbf{s})=V_{IB}(\mathbf{k})V_{IB}(-\mathbf{s})$ is not valid in general. 
To continue studying the beyond-Fröhlich corrections, $c$ is set to zero. There should not be a meaningful difference between the polaronic energy for $c=0$ and $c=0.0023$ \cite{privatecomm}. In that case, the potential reduces to an exponential in momentum space, and the condition holds. 
The remainder of the derivation of the effective action is identical to that in Ref. \cite{ichmoukhamedov2019feynman} except that the contact interaction potential is replaced with the long-range ion-atom interaction potential.

\begin{figure}
    \centering
    \includegraphics[scale=1]{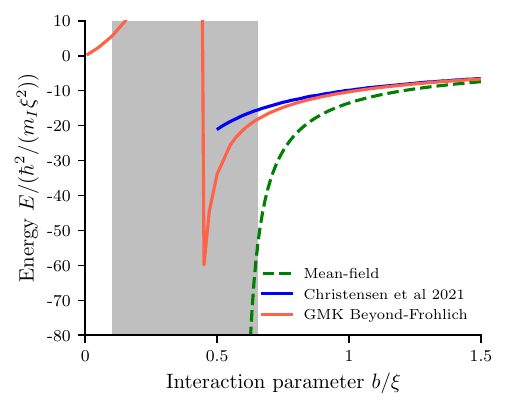}
    \caption{The polaronic energy of the charged Bose polaron using the beyond-Fröhlich Hamiltonian as a function of the interaction parameter $b$ at zero temperature with $m_B=m_I$, $c=0$, $a=\hbar^2\xi^2/m_B$, and $n_0\xi^3=1$. The red line indicates the result using the GMK approach, while the green line indicates the mean-field energy. For large $b$ they converge to the same value. There is a divergence around $b=0.45\xi$ with the GMK result. However, the Bogoliubov approximation fails in this region (grey patch), and no conclusions can be made.}
    \label{fig2}
\end{figure}

The effective action at zero temperature, excluding the kinetic energy term, is given by
\begin{align}
    \tilde{\mathcal{S}}= & -\hbar\sum_{n=0}^\infty \left(\frac{\pi^2a}{b}\right)^n \frac{\pi^4a^2n_0}{\hbar(2V\hbar)^{n+1}b^2}\int_0^{\infty}d\tau_1...\int_0^{\infty}d\tau_{n+2} \nonumber \\
    & \times \sum_{\mathbf{k}_1,...,\mathbf{k}_{n+1}}\prod_{j=1}^{n+1}V^2_{\mathbf{k}_j}e^{-2k_jb-\omega_{k_j}|\tau_{j+1}-\tau_j|}e^{i\mathbf{k}_j\cdot(\mathbf{r}(\tau_{j+1})-\mathbf{r}(\tau_j))},
\end{align}
The following step is to calculate the expectation values of the actions. 
Compared to \cite{ichmoukhamedov2019feynman}, the more general GMK trial action is used instead of the Feynman trial action. The expectation value and the GMK trial action's free energy are known and given in \cite{ichmoukhamedov2022general}. 
The expectation value of the effective action is calculated in the same way as in Refs.~\cite{ichmoukhamedov2019feynman,ichmoukhamedov2022general} by using the random phase approximation (RPA) to split the expectation value of products of the impurity density. 
The resulting equation for the free energy is
\begin{align}
    F \leq & \frac{3\hbar}{2\pi}\int_0^\infty d\nu \left[\ln\left(1+\frac{x(\nu)}{\nu^2}\right)-\frac{x(\nu)}{x(\nu)+\nu^2}\right] \nonumber \\
    & -\frac{\pi^2an_0/b}{1-\frac{\pi^2a}{\hbar Vb}\sum_{\mathbf{k}}V^2_{\mathbf{k}}e^{-2kb}\int_0^{\infty}due^{-\omega_ku}\mathcal{F}_{\mathbf{k}}(u)}.
\end{align}
The Fr\"ohlich result can be derived by expanding the last term for small values of $a/b$. The equation for the kernel needs to be calculated in a similar way as for the Fr\"ohlich case \cite{ichmoukhamedov2022general} and is given by
\begin{align}
    x(\nu) & = \frac{4\pi^4a^2n_0}{3\hbar Vb^2m_I} \nonumber \\
    &\times \frac{\sum_{\mathbf{k}}k^2V^2_{\mathbf{k}}e^{-2kb}\int_0^{\infty}due^{-\omega_ku}\mathcal{F}_{\mathbf{k}}(u)\sin^2(\nu u/2)}{(1-\frac{\pi^2a}{\hbar Vb}\sum_{\mathbf{k}}V^2_{\mathbf{k}}e^{-2kb}\int_0^{\infty}due^{-\omega_ku}\mathcal{F}_{\mathbf{k}}(u))^2}.
\end{align}
The result for the energy is plotted in Figure \ref{fig2} for $c=0,a=\hbar^2\xi^2/m_B,n_0\xi^3=1$, and $m_B=m_I$. The energy decreases in the attractive polaron regime as $b$ becomes smaller. For large $b$, the energy converges to the mean-field result $\pi^2n_0a_{IB}/b$ where $a_{IB}$ is the ion-atom scattering length calculated in Ref. \cite{christensen2021charged} by solving the Schrödinger equation. 
The result from Ref. \cite{christensen2021charged} using the ladder approximation and the Bethe-Salpeter equation is also plotted. 
There is a divergence around $b=0.45\xi$ where the repulsive polaron regime is entered, and the energy decreases again as $b$ becomes smaller. 
The Bogoliubov approximation is not valid for the value of $b$ of the divergence, and the results are only correct for $b\gtrsim 0.65$ and $b\lesssim 0.1$. 
The result from Ref. \cite{christensen2021charged} is also based on the Bogoliubov approximation and is also not expected to be valid anymore in that region. It can be seen that the energy calculated using the path-integral method is comparable and even better than the field-theory result from Ref. \cite{christensen2021charged}.

\section{Validity of the Bogoliubov approximation}
\begin{figure}
    \centering
    \includegraphics[scale=1]{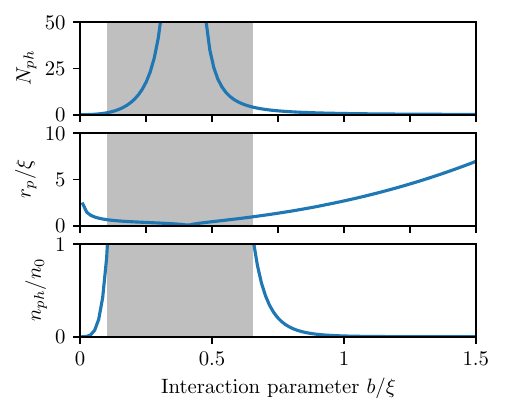}
    \caption{The number of phonons (top), the polaron radius (middle), and the Bogoliubov ratio (bottom) calculated using the GMK Beyond-Fröhlich approach as a function of the ion-atom interaction parameter $b$ for $c=0,a=\hbar^2\xi^2/m_B,n_0\xi^3=1$, and $m_B=m_I$. The grey area indicates the region where the Bogoliubov approximation breaks down and the Bogoliubov ratio becomes larger than one.}
    \label{fig4}
\end{figure}

The Bogoliubov approximation on which our results are based is only valid when the number of phonons and depletion is small enough close to the impurity, $n_{ph}/n_0 \ll 1$. The density of phonons locally for the polaron can be estimated as follows
\begin{equation}
    n_{ph} = \frac{N_{ph}}{V_p} = \frac{N_{ph}}{4\pi r_p^3/3},
\end{equation}
with $V_p=4\pi r_p^3/3$ the volume of the polaron with $r_p$ is the polaron radius. The number of phonons can be related to the derivative of the energy by the use of the Feynman-Hellmann theorem \cite{peeters1985radius}
\begin{equation}
    N_{ph} = \sum_{\mathbf{k}}\langle \hat{\alpha}^\dagger_{\mathbf{k}}\hat{\alpha}_{\mathbf{k}}\rangle = \sum_{\mathbf{k}}\left\langle\frac{\partial H}{\partial \hbar\omega_k}\right\rangle=\sum_{\mathbf{k}}\frac{\partial E}{\partial \hbar\omega_k}.
\end{equation}
The equation for the energy, including beyond-Fröhlich corrections from the previous section, can be used, resulting in the following expression for the number of phonons at zero temperature
\begin{equation}
    N_{\mathbf{k}}=\frac{\frac{\pi^4a^2n_0}{V\hbar^2 b^2}\int_0^{\infty}duuV^2_{\mathbf{k}}e^{-2kb-\omega_{k}u}\mathcal{F}_k(u)}{\left(1-\sum_{\mathbf{q}}\frac{\pi^2a}{bV\hbar}\int_0^{\infty}duV^2_{\mathbf{q}}e^{-2qb-\omega_{q}u}\mathcal{F}_q(u)\right)^2}.
\end{equation}
By comparing the density matrix of the trial action \cite{ichmoukhamedov2021path} to the density matrix of the harmonic oscillator, a polaron radius can be defined similarly to the original Feynman action polaron radius \cite{tempere2009feynman}. This was done in \cite{houtput2022voorbij} for the anharmonic solid-state polaron problem. The polaron radius is given by
\begin{equation}
    r_p=\sqrt{\frac{3\pi\hbar}{4m_I}\left(\int_0^\infty d\nu \frac{x(\nu)}{x(\nu)+\nu^2}\right)^{-1}}.
\end{equation}
The phonon number, polaron radius, and the Bogoliubov ratio $n_{ph}/n_0$ are shown in Figure \ref{fig4} for different values of $b$ with $m_B=m_I, c=0, a=\hbar^2\xi^2/m_B$ and $n_0\xi^3=1$. The number of phonons increases the smaller $b$ is until the divergence around $0.45\xi$. The number of phonons then decreases for smaller $b$. The polaron radius does not have a divergence and decreases as $b$ becomes smaller as expected till the repulsive polaron regime is entered. For $b \lesssim 0.65\xi$, the Bogoliubov ratio becomes larger than one, and the Bogoliubov approximation fails. The energy, phonon number, and polaron radius results become inaccurate. However, for small values of $b$ around $0.1\xi$ in the repulsive polaron regime, the Bogoliubov approximation is valid again.
\section{\label{sec:optabs}Weak-coupling optical absorption}
In the previous sections, the ground state energy of the charged Bose polaron was calculated. In this section, another important quantity, the optical absorption, will be studied. 
For the solid-state polaron, the optical absorption shows a typical polaron band that is a key experimental signature revealing the presence of polarons. 
It has yet to be calculated for an (ionic) impurity in a Bose-Einstein condensate. 
So far only methods suited for neutral impurities have been considered to detect and study ions in Bose condensates. 
We propose direct absorption, made possible by the fact that the impurities are charged, as an efficient and direct alternative method to study ions in condensates.

\begin{figure*}
    \centering
    \includegraphics[scale=1]{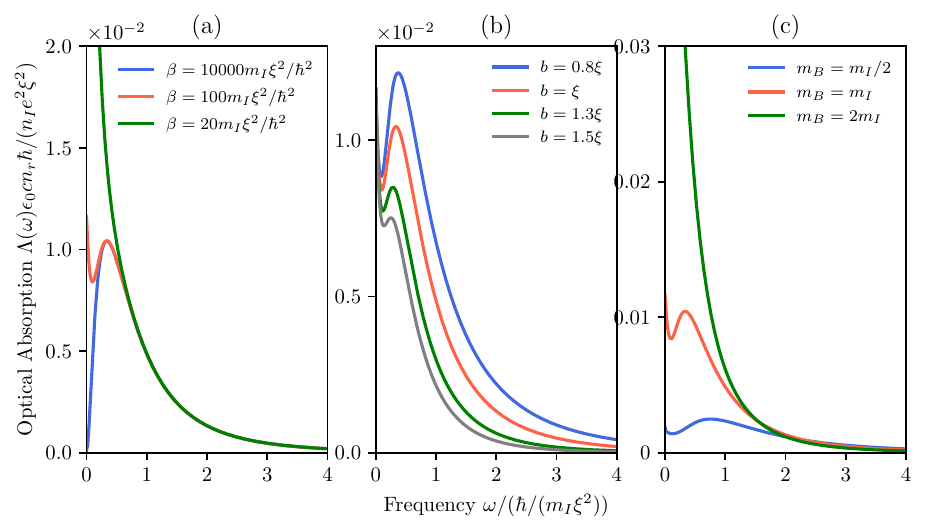}
    \caption{The optical absorption for a charged Bose polaron with $c=0.0023\xi$, $\alpha=1$, and $R=\xi$. (a) shows the temperature dependence for fixed $b=\xi$ and $m_B=m_I$, (b) shows the dependence on the potential depth $b$ for $\beta=100m_I\xi^2/\hbar^2$ and $m_B=m_I$, and (c) shows the dependence on the boson mass for $\beta=100m_I\xi^2/\hbar^2$ and $b=\xi$.}
    \label{fig3}
\end{figure*}

The method used in this section to calculate the optical absorption is based on Ref.~\cite{houtput2022optical}, where it was calculated in the weak-coupling regime for a solid-state (anharmonic) polaron. The optical absorption $\Lambda(\omega)$ is related to the real part of the optical conductivity via $\Lambda(\omega)\epsilon_0cn_r = \Re[\sigma(\omega)]$ with $\epsilon_0$ the vacuum permittivity, $c$ the speed of light, and $n_r$ the refractive index of the medium. The optical conductivity can be calculated using the Kubo formalism
\begin{equation}
    \sigma(\omega)=\frac{in_Ie^2}{m_B\omega}+\frac{1}{V\hbar\omega}\int_0^\infty e^{i\omega t}\langle[J_x(t),J_x(0)]\rangle dt,
\end{equation}
with $n_I$ the impurity density.
The optical conductivity is thus related to the current-current correlation function. Ref.~\cite{houtput2022optical} is followed where the memory function formalism is used. 
The optical conductivity can be rewritten in terms of the memory function $\Sigma(\omega)$ as
\begin{equation}
     \sigma(\omega)=\frac{in_Ie^2}{m_I}\frac{1}{\omega-\Sigma(\omega)}.
\end{equation}
Here $\Sigma(\omega)$ is related to the so-called approximate memory function $\Sigma_0(\omega)$ via
\begin{equation}
     \Sigma(\omega)=\frac{\Sigma_0(\omega)}{1+\frac{\Sigma_0(\omega)-\Sigma_0(0)}{\omega}}.
\end{equation}
The approximate memory function in turn can be written in terms of the force-force correlation function
\begin{equation}
    \Sigma_0(\omega)=\frac{1}{im_I\omega}\int_0^\infty \left(e^{i\omega t}-1\right)\langle[F_x(t),F_x(0)]\rangle dt,
\end{equation}
with $F_x$ the force operator defined as $F_x(t) = dP_x/dt=i[H,P_x]/\hbar$.
The weak coupling requirement is because the expectation values were factorized in an expectation value of the phonon operators and an expectation value of the electron operators, and this only holds up to first order in the coupling strength $\alpha$. So, it is assumed that there is no phonon entanglement like in the Lee-Low-Pines approach.
As the weak-coupling limit is considered, the beyond-Fröhlich terms do not have to be accounted for. The weak-coupling requirement can be checked by comparing the Lee-Low-Pines result for the energy with the general memory kernel result and seeing if they are similar. The real and imaginary parts of $\Sigma_0(\omega)$ are
\begin{equation}
    \Re[\Sigma_0(\omega)]=\frac{2\omega}{\pi}\int_0^\infty \frac{\Im[\Sigma_0(\omega')]-\Im[\Sigma_0(\omega)]}{\omega'^2-\omega^2}d\omega',
\end{equation}
and
\begin{align}
    \Im[\Sigma_0(\omega)]= & -\frac{\pi}{3m_I\hbar\omega}\sum_{\mathbf{k}}k^2\int_{-\infty}^\infty [ 1+n_B(\omega') \nonumber \\
    & +n_B(\omega-\omega') ] S(\mathbf{k},\omega-\omega')M(\mathbf{k},\omega')d\omega'.
\end{align}
Here $n_B(\omega)=1/(\exp(\hbar\beta\omega)-1)$ is the Bose-Einstein distribution, and $S(\mathbf{k},\omega)$ and $M(\mathbf{k},\omega)$ are the dynamic structure factor of the impurity and the phonon spectral function, respectively, given by
\begin{equation}
    S(\mathbf{k},\omega)=\frac{1}{2\pi}(1-e^{-\hbar\beta\omega})\int_{-\infty}^\infty \langle\hat{\rho}_I(\mathbf{k},t)\hat{\rho}_I(-\mathbf{k},0)\rangle_0e^{i\omega t}dt,
\end{equation}
and
\begin{equation}
    M(\mathbf{k},\omega)=\frac{1}{2\pi}(1-e^{-\hbar\beta\omega})\int_{-\infty}^\infty \langle\hat{\mathcal{F}}(-\mathbf{k},t)\hat{\mathcal{F}}(\mathbf{k},0)\rangle_0e^{i\omega t}dt,
\end{equation}
where
\begin{equation}
    \hat{\mathcal{F}}(\mathbf{k})=\sqrt{N_0}V_{IB}(\mathbf{k})V_{\mathbf{k}}(\hat{\alpha}^\dagger_{\mathbf{k}}+\hat{\alpha}_{-\mathbf{k}}).
\end{equation}
Only a single ionic impurity is studied in this paper, so the dynamic structure factor is a delta function $\delta(\omega-\hbar k^2/(2m_I))$. The phonon spectral function has been calculated for a Fröhlich-like Hamiltonian in the literature and is given by \cite{houtput2022optical}
\begin{equation}
    M(\mathbf{k},\omega)=N_0V^2_{IB}(\mathbf{k})V^2_{\mathbf{k}}(\delta(\omega-\omega_{\mathbf{k}})-\delta(\omega+\omega_{\mathbf{k}})).
\end{equation}
The optical absorption is calculated for small coupling strength $\alpha=1$, $c=0.0023\xi$, and $R=\xi$ for different values of potential depth $b$, boson mass $m_B$, and temperature $\beta$. For the values used, the difference between the Lee-Low-Pines energy and the general memory kernel energy is less than one percent. The weak-coupling condition is satisfied. Figure \ref{fig3}(a) shows the optical absorption for $b=\xi$ and $m_B=m_I$. For zero temperature (blue line), there is a single so-called polaron peak at a certain frequency value. Compared to the solid-state polaron results \cite{houtput2022optical}, the optical absorption does not become non-zero when the frequency is larger than the longitudinal-optical phonon frequency, but due to the dispersion, the optical absorption is finite everywhere except at zero frequency. The effect of finite temperature is similar to the solid-state polaron case; the optical absorption becomes non-zero at zero frequency. For large enough temperatures, the polaron peak is washed out due to broadening the "peak" around zero frequency. 
The effect of the depth of the potential $b$ is that the larger $b$ becomes, the closer the polaron peak is to zero frequency and the smaller the peak height (and it will become more washed out at finite temperature). 
This can be seen in Figure \ref{fig3}(b) for fixed $\beta=100m_I\xi^2/\hbar^2$ and $m_B=m_I$. As can be observed in Figure \ref{fig3}(c) for $\beta=100m_I\xi^2/\hbar^2$ and $b=\xi$, the heavier the impurity, the further the polaron peak is from zero frequency and the smaller the height. 

Considering a typical \ce{^{87}_{} Rb} Bose-Einstein condensate of $N=10^5$ atoms with scattering length $a_s=100a_0$ and healing length $\xi=270$ nm, the frequency of the polaron peak and thus the frequencies of the incoming photons is around \SI{3}{\kilo\hertz}. 
The power $P$ dissipated by the ion is related to the memory function and the intensity $I_{\textrm{avg}}$ of the incoming radiation by
\begin{equation}
    P = I_{\textrm{avg}}\frac{2e^2}{m_I\epsilon_0c}\Re\left(\frac{i}{\omega-\Sigma(\omega)}\right).
\end{equation}
Using the parameters from Figure \ref{fig3}(a), the absorption cross-section of the ion (i.e. the ratio of the dissipated power to the intensity of the incoming beam) is \SI{447}{\nano\meter\squared}. For a RF antenna delivering \SI{10} {\micro\watt\per\meter\squared} on resonance with the absorption peak, this leads to an experimentally detectable heating rate of the order of 1-\SI{10}{\nano\kelvin\per\second} for $10^5$ atoms in a parabolic trap. This scales linearly with the number of ions, the coupling strength (at zero temperature), and the antenna power, so that experimenters can tune the heating rate to bring it in their desired regime for detection.

\section{\label{sec:conclusion}Conclusion and Outlook}
Using the Feynman variational path-integral approach with a general memory kernel, we studied the ground state energy of a charged impurity in a Bose-Einstein condensate within the Bogoliubov approximation. The results were compared to that of the Lee-Low-Pines weak coupling approach and to the result of a T-matrix calculation in the ladder approximation \cite{christensen2021charged}. 
It was found that the general memory kernel results in a minor improvement over the original Feynman approach \cite{casteels2011polaronic} for the Fr\"ohlich Hamiltonian, which indicates that the general kernel is only necessary when a contact interaction potential is used. 
Including beyond-Fr\"ohlich corrections results in the emergence of a divergence in the polaronic energy indicating a transition between the repulsive and attractive polaron regime. 
Also, the path-integral approach with the beyond-Fr\"ohlich Hamiltonian gives a improved result compared to the result in Ref.~\cite{christensen2021charged} which is based on a field-theory calculation. The validity of the Bogoliubov approximation was also studied within the path-integral formalism. 

Charged impurities can be probed by direct coupling to the electromagnetic field, resulting in a polaron peak in the optical absorption. We calculated the optical absorption for an ionic impurity in a condensate in the weak ion-atom and atom-atom interaction regime within the Kubo formalism. The effect of temperature and other parameters was studied. The polaron peak in the absorption remains a robust feature also at finite temperatures. Our results show that measuring the optical absorption of charged impurities in condensates provides a direct way to study the ionic polarons. It is complementary to currently used methods of injection (and ejection) spectroscopy, in that it can be used as a probe to study polarons at arbitrary times after their formation without removing or injecting the impurity.

Calculating the optical absorption for intermediate coupling strengths within the Bogoliubov approximation using the Feynman-Hellwarth-Iddings-Platzmann method \cite{feynman1962mobility,devreese1972optical} is a possible future research topic. Another interesting question is how the energy and optical absorption of multiple ionic impurities would look as there is a Coulomb repulsion between the ions compared to the case of neutral impurities \cite{casteels2013bipolarons}. Experimentally, it would be interesting to measure the optical absorption of single and multiple ions in a condensate and compare it with the theoretical results.

\begin{acknowledgments}
We gratefully acknowledge fruitful discussions with G. Bruun, T. Ichmoukhamedov, and M. Houtput. We also thank E.R. Christensen and A. Camacho-Guardian for sharing their results of the ground-state energy from \cite{christensen2021charged}. We acknowledge financial support by the Research Foundation Flanders (FWO), Projects No. GOH1122N, No. G061820N, and No. G060820N, and by the University Research Fund (BOF) of the University of Antwerp.
\end{acknowledgments}

%
%
%

\end{document}